\documentclass[fleqn,twoside]{article}
\usepackage[headings]{espcrc2}

\usepackage{amssymb} \usepackage{amsmath}
 \usepackage{epsfig} \usepackage{pifont}
\def\agoth{\relax\ifmmode{\mathfrak A}\else{$\,{\mathfrak A}${ }}\fi}
\def\acal{\relax\ifmmode{\mathcal{A}}\else{${\mathcal{A}}${ }}\fi}
\def\MSbar{\relax\ifmmode\overline{\rm MS}\else{$\overline{\rm MS}${ }}\fi}

\readRCS
$Id: espcrc2.tex,v 1.2 2004/02/24 11:22:11 spepping Exp $
\usepackage[figuresright]{rotating}

\title{A modification of minimal analytic QCD at low energies}

\author{G. Cveti\v c\address[MCSD]{Physics Department, 
Universidad T\'ecnica Federico Santa Mar\'{\i}a, Valpara\'{\i}so, Chile},
        C. Valenzuela\addressmark,
and I. Schmidt\addressmark\thanks{The work supported in 
part by Fondecyt grants No.~1050512 (G.C.), 1030355 (I.S.) and 
Mecesup grant USA0108 (C.V.).
Presented at QCD05, 4-8 July 2005, Montpellier, France.}}
       
\begin{document}

\begin{abstract}
We present an approach which allows for modifications of the (minimal)
analytic QCD model of Shirkov, Solovtsov {\it et al.\/}. 
The discontinuity function of the (minimal) analytic QCD coupling 
parameter is changed at the low time-like momenta by a simple 
ansatz motivated by the vector meson dominance for the vector channel 
of the $e^+e^-$ hadronic decay ratio $R_V(s)$. In this way, the additional 
low energy parameters introduced in the coupling parameter make it possible
to obtain the correct value of the semihadronic tau decay ratio
without having to introduce the unusually heavy quark masses of
$u$, $d$ and $s$ quarks, in contrast to the minimal analytic QCD. 
Some numerical results are presented in the
leading skeleton approximation for a simple version of such a model.
We hope that a more systematic analysis of such models leads
to an approximately universal coupling parameter.
\vspace{1pc}
\end{abstract}

\maketitle


It has been well known that the usual perturbative QCD coupling
$a_{\rm pt}(Q^2) \equiv \alpha_s(Q^2)/\pi$ 
(where: $Q^2\!=\!- q^2\!=\!- (q^0)^2\!+\!{\vec q}^2$)
has nonphysical singularities outside the time-like
momentum region, specifically a Landau cut on the
space-like axis: $0 < Q^2 \leq \Lambda^2$. This can be
seen, for example, from the expansion
\begin{eqnarray}
a_{\rm pt}(Q^2) &=& \frac{1}{\beta_0 \ln (Q^2/\Lambda^2)} 
- \frac{\beta_1}{\beta_0^3} 
\frac{ \ln \ln (Q^2/\Lambda^2)}{\ln^2 (Q^2/\Lambda^2) }
\nonumber\\
&&+ {\cal O}(1/\ln^3(Q^2/\Lambda^2) \ .
\label{aptexp}
\end{eqnarray}
Direct application of the Cauchy theorem gives
the following dispersion relation for $a_{\rm pt}$:
\begin{equation}
a_{\rm pt}(Q^2) = \frac{1}{\pi} \int_{\sigma= - \Lambda^2}^{\infty}
\frac{d \sigma \rho_1(\sigma) }{(\sigma + Q^2)} \ ,
\label{aptdisp}
\end{equation}
where $\rho_1(\sigma)$ is the (pQCD) discontinuity function of $a_{\rm pt}$
along the cut axis in the complex $Q^2$-plane:
$\rho_1(\sigma) = {\rm Im} a_{\rm pt}(-\sigma - i \epsilon)$ .

The physical observables are, by special relativity and
causality, analytic functions of the associated physical
momentum squared $q^2 \equiv - Q^2$ in the entire complex $Q^2$
plane with the exception of the time-like axis ($Q^2 < 0$).
Any QCD observable is a function of the invariant coupling
$a(Q^2)$, so both should have the same analyticity properties.
Therefore, the cut sector $0 < Q^2 \leq \Lambda^2$ in
$a_{\rm pt}(Q^2)$, Eqs.~(\ref{aptexp})-(\ref{aptdisp}),
is nonphysical. A simple analytization, which is minimal
in a sense, is obtained by eliminating the unphysical cut
in the dispersion relation (\ref{aptdisp}) while keeping 
elsewhere the same discontinuity function $\rho_1(\sigma)$, 
thus leading to a specific analytic coupling $\acal_1(Q^2)$
\cite{ShS}
\begin{equation}
\acal_1(Q^2) = \frac{1}{\pi} \int_{\sigma= 0}^{\infty}
\frac{d \sigma \rho_1(\sigma) }{(\sigma + Q^2)} \ .
\label{madisp}
\end{equation}
Calculational aspects of this prescription were investigated in
\cite{Magradze}.
Other approaches to analytization of $a_{\rm pt}$
have focused on the analyticity properties of the
beta function \cite{Nesterenko,Raczka}, or the addition of
power correction terms $1/(Q^2)^n$
to $\acal_1$ \cite{Alekseev}. Our approach
considers a modification of the pQCD discontinuity
function $\rho_1(\sigma)$ of Eq.~(\ref{madisp})
in the $\sigma \sim \Lambda$ region of the time-like axis,
resulting in a different analytic coupling ${\overline \acal}_1(Q^2)$
\begin{equation}
{\overline \acal}_1(Q^2) = \frac{1}{\pi} \int_{\sigma= 0}^{\infty}
\frac{d \sigma {\overline \rho}_1(\sigma) }{(\sigma + Q^2)} \ .
\label{adisp}
\end{equation}
We recall that relation (\ref{adisp})
defines the coupling in the entire complex
$Q^2$-plane except on the time-like axis $Q^2 < 0$.
Then, in the analytic approach the
coupling on the time-like axis $s = - Q^2 > 0$
can be defined in the following convenient way \cite{time-like}
\begin{equation}
{\overline \agoth}_1(s) = \frac{i}{2 \pi} 
\int_{- s + i \epsilon}^{-s - i \epsilon} \frac{d \sigma^{\prime}}{\sigma^{\prime}}
{\overline \acal}_1(\sigma^{\prime}) \ ,
\label{atime-like}
\end{equation}
where the integration is in the $Q^2 \equiv \sigma^{\prime}$ plane avoiding
the (time-like) cut $\sigma^{\prime} < 0$. This relation between
${\overline \acal}_1(Q^2)$ and ${\overline \agoth}_1(s)$ is the same
as the one between the (vector channel) Adler function $D_V(Q^2)$ 
and the $e^+e^-$ hadronic scattering cross section ratio $R_V(s)$.
In addition to relations (\ref{adisp}) and (\ref{atime-like}),
the following relations hold \cite{Sh}:
\begin{eqnarray}
{\overline \agoth}_1(s) &=& \frac{1}{\pi} 
\int_s^{\infty} \frac{d \sigma}{\sigma}
{\overline \rho}_1(\sigma) \ ,
\label{adispH}
\\
{\overline \acal}_1(Q^2) & = & Q^2 \int_0^{\infty} 
\frac{ ds {\overline \agoth}_1(s) }{(s + Q^2)^2} \ ,
\label{aspace-like}
\\
\frac{d}{d \ln \sigma} {\overline \agoth}_1(\sigma) &=&
- \frac{1}{\pi} {\overline \rho}_1(\sigma) \ .
\label{Hrho}
\end{eqnarray}
The value of the coupling parameter at the origin is in general finite
(${\overline \acal}(0) = {\overline \agoth}_1(0) = 
(1/\pi) \int_0^{\infty} d \sigma {\overline \rho}_1(\sigma)/{\sigma}$.)

The minimal analytic QCD model \cite{ShS,time-like} [cf.~Eq.~(\ref{madisp})]
contains only one free parameter -- the scale $\Lambda$ contained
in the pQCD discontinuity function $\rho_1(\sigma)$. By fitting the
measured values of high energy QCD observables 
(at $Q$ or ${\sqrt{s}} \stackrel{>}{\sim} 10$ GeV) to the model,
the value $\Lambda({\overline {\rm MS}})_{n_f=5} \approx 0.26$ GeV 
was obtained \cite{Sh}, corresponding to 
$\Lambda({\overline {\rm MS}})_{n_f=3} \approx 0.4$ GeV.
At lower energies, however, a problem appeared.
The $V$-channel of the semihadronic $\tau$ decay ratio,
without the strangeness production, has been measured
with high precision by the ALEPH and OPAL \cite{ALEPH}
Collaborations with the result: 
$R_{\tau,V}(\triangle S\!=\!0) \approx 1.19$,
and where this quantity has been normalized so that
in pQCD: $R_{\tau,V}(\triangle S\!=\!0) = [ 1 +
a_{\rm pt} + {\cal O}(a^2_{\rm pt}) ]$.
However, fitting this value with the prediction of the
minimal analytic QCD gave: 
$\Lambda({\overline {\rm MS}})_{n_f=3} \approx 0.8$ GeV
\cite{MSSY}. This problem was then solved by the same authors
in Ref.~\cite{MSS}, where the value 
$\Lambda({\overline {\rm MS}})_{n_f=3} \approx 0.4$ GeV
was obtained at the price of introducing large quark
masses $m_u \approx m_d \approx 0.25$ GeV which implied
very strong threshold effects in the quantity
$R_{\tau,V}(\triangle S\!=\!0)$.  
The introduction of (effective) heavy quark masses may be 
regarded as unattractive or hard to justify theoretically.

A different possibility would be to include (some)
nonperturbative effects {\it within} the time-like
coupling $\agoth_1(s)$ [$\Leftrightarrow \acal_1(Q^2)$].
For that purpose let us 
consider first the isovector hadronic spectral function
$R_V(s)$ [with the normalization
$R_V(s) = 1 + a_{\rm pt} + {\cal O}(a^2)$]
which has been measured with high precision \cite{ALEPH}. 
At low energies  $\sqrt{s} \stackrel{<}{\sim} 1$ GeV 
it is dominated by the $\rho$-meson resonance 
($M_{\rho} = 0.776$ GeV). On the other hand,
the Adler function is obtained by the relation
\begin{eqnarray}
D_V(Q^2) & = & Q^2 \int_0^{\infty} 
\frac{ ds R_V(s) }{(s + Q^2)^2} \ ,
\label{DVRV}
\end{eqnarray}
which is analogous to Eq.~(\ref{aspace-like}).
If in the aforementioned resonance region we use a
narrow width approximation (NWA) for $R_V(s)$,
in the spirit of the Vecton Meson Dominance (VMD), then
\begin{eqnarray}
\lefteqn{
(4 \pi^2)^{-1} R_V(s)_{\rm VMD} =
2 f_{\rho}^2 M_{\rho}^2 \delta(s\!-\!M_{\rho}^2)
}
\nonumber\\
&& + (4 \pi^2)^{-1} R_V(s)_{\rm pQCD} 
\theta(s\!-\!M_{\rm cut}^2) \ ,
\label{RVVMD}
\end{eqnarray}
where $f_{\rho}^2 = (f_{\rho}^2)_{\rm NWA} \approx 0.0305$,
and $\theta(x)$ is the step function ($+1$ for $x > 0$,
zero for $x < 0$).
The cut value $M_{\rm cut}^2 \approx 1.53 \ {\rm GeV}^2$ 
\cite{PPR} was fixed so that the corresponding Adler function
(\ref{DVRV}) is closest to the ``experimental'' Adler
function $D_V(Q^2)_{\rm exp}$
[obtained via relation (\ref{DVRV}) from the
measured $R_V(s)_{\rm exp}$] in the low $Q^2$ sector
($0\!\leq\!Q^2\!< 2 \ {\rm GeV}^2$) where the quark masses can
presumably be neglected. The agreement of the two Adler
functions is then remarkably close there.
Since in the analytic QCD 
$R_V(s) = 1 + {\overline \agoth}_1(s)
+ {\cal O}(({{\overline \agoth}_1})^2)$, then
the VMD form (\ref{RVVMD}) suggests
\begin{eqnarray}
\lefteqn{
{\overline \agoth}_1(s) =  \left[ R_V(s) - 1 \right] + 
 {\cal O}(({{\overline \agoth}_1})^2) }
\nonumber\\
& \approx & 
\left[ c_f M_r^2 \delta(s\!-\!M_r^2) - 1 \right]
\nonumber\\
&&\!\!\!\!\!\!\!\!\! 
+ \left( \agoth_1(s) + 1 \right) \theta(s\!-\!M_0^2) +
{\cal O}(({{\overline \agoth}_1})^2) ,
\label{VMD1}
\end{eqnarray}
where $\agoth_1(s)$ is the minimal analytic QCD
time-like coupling (containing the parameter $\Lambda$);
$c_f$, $M_r$ and $M_0$ are additional parameters
whose values are
comparable to $8 \pi^2 f^2_{\rho}$, $M_{\rho}$ and
$M_{\rm cut}$ of Eq.~(\ref{RVVMD}), respectively.
Neglecting higher order terms in (\ref{VMD1}) gives
\begin{eqnarray}
{\overline \agoth}_1(s) =  
\left\{ 
\begin{array}{cc}
\left[ c_f M_r^2 \delta(s\!-\!M_r^2)\!-\!1 \right] &
(s\!<\!M_0^2)
\\
\agoth_1(s) & (s\!\geq\!M_0^2)
\end{array}
\right\}
\label{atime-like2}
\end{eqnarray}
and $M_r < M_0$ is implicitly assumed. Applying the transformation
(\ref{aspace-like}) to this VMD-motivated time-like coupling,
the corresponding analytic space-like coupling is obtained
\begin{eqnarray}
{\overline \acal}_1(Q^2) &=&
c_f \frac{M_r^2 Q^2}{ (Q^2 + M_r^2)^2} -
d_f  \frac{M_0^2}{(Q^2 + M_0^2)}
\nonumber\\
&& + \frac{1}{\pi} \int_{\sigma= M_0^2}^{\infty}
\frac{ d \sigma \rho_1(\sigma) }{(\sigma + Q^2) } \ ,
\label{aspace-like2}
\end{eqnarray}
where
\begin{equation}
d_f = 1 + \frac{1}{\pi} \int_{\sigma= M_0^2}^{\infty}
\frac{d \sigma}{\sigma} \rho_1(\sigma) \ .
\label{df}
\end{equation}
The presented version of the model contains
four parameters: $c_f$, $M_r$, $M_0$, and $\Lambda$
(the latter is contained in $\agoth_1(s)$ and in $\rho_1(s)$).
In general, analytic coupling (\ref{aspace-like2})
differs from the minimal analytic coupling $\acal_1(Q^2)$,
Eq.~(\ref{madisp}), by terms $\sim\!(\Lambda^2/Q^2)$.

In this presentation we choose, for simplicity, the
additional condition that the two
couplings virtually merge at high momenta, i.e., that
${\overline \acal}_1(Q^2) - \acal_1(Q^2) \sim (\Lambda^2/Q^2)^2$.
Then, (a) the high energy QCD observables 
($Q, \sqrt{s} \stackrel{>}{\sim} 10$ GeV) give the value
$\Lambda_{n_f\!=\!3} \approx 0.4$ GeV as in the minimal
analytic model \cite{Sh}, and (b)
an additional relation is obtained
\begin{eqnarray}
\lefteqn{
M_r^2 c_f = \frac{1}{\pi} \int_0^{M_0^2} d \sigma \rho_1(\sigma) +
}
\nonumber\\
&& + M_0^2 \left[ \frac{1}{\pi} \int_{M_0^2}^{\infty}
\frac{d \sigma}{\sigma} \rho_1(\sigma) + 1 \right].
\label{cfrel}
\end{eqnarray}
Therefore, in such a case the model contains two 
yet undetermined parameters, the scales $M_r$ and $M_0$,
i.e.,
${\overline \acal}_1 = {\overline \acal}_1(Q^2; M_r, M_0)$,
${\overline \agoth}_1 = {\overline \agoth}_1(s; M_r, M_0)$.
These two parameters can then be fixed, for example, 
by the condition
that the measured semihadronic decay value 
$R_{\tau,V}(\triangle S\!=\!0)$ and a ``measured''
value of the $V$-channel Adler function $D_V(Q^2)$
(at a given chosen low $Q^2$) be  reproduced in the model.
In this presentation, we will use the approximation
of the leading skeleton (l.s.), also known
as large-$\beta_0$ resummation. According to 
this resummation \cite{Neubert1,Neubert2}
\begin{eqnarray}
\lefteqn{
D_V(Q^2)^{\rm (l.s.)} =
}  
\nonumber\\
&&
1 +
\frac{1}{4} \int_0^{\infty} \!\! dt w_D(t) 
a_{\rm pt}(t Q^2 e^{-5/3}),
\label{DVsl1}
\\
\lefteqn{
R_{\tau,V}(\triangle S\!=\!0)^{\rm (l.s.)} = 
} 
\nonumber\\
&&
1 +
\frac{1}{4} \int_0^{\infty} \!\!\! dt W_{\tau}(t) 
a_{\rm pt}^{\rm (t.l.)}(t m_{\tau}^2 e^{-5/3}),
\label{Rtausl1}
\end{eqnarray}
where the scheme invariant distribution
functions $t w_D(t)$ and $t W_{\tau}(t)$
are explicitly known \cite{Neubert1,Neubert2} 
(cf.~also \cite{BBB})
and are peaked at around $t \sim 1$. The factor
$e^{-5/3}$ in the above formulas means that the
${\overline {\rm MS}}$ scheme is used for $a_{\rm pt}$,
i.e., $\Lambda_{\overline {\rm MS}}$ appears there.
In Eq.~(\ref{Rtausl1}), $a_{\rm pt}^{\rm (t.l.)}$
is a time-like coupling \cite{Neubert2}.
The expressions (\ref{DVsl1}) and (\ref{Rtausl1})
have renormalon ambiguities and the
time-like analytic continuation ambiguities in pQCD,
both manifested via the existence of unphysical
singularities of $a_{\rm pt}$ on the Landau cut.
In our model, there is no Landau cut, thus no
ambiguities, and $a_{\rm pt}$
in the above expressions gets replaced by the
space-like and the time-like couplings, respectively
\begin{eqnarray}
\lefteqn{
D_V(Q^2)^{\rm (l.s.)} = 
}
\nonumber\\
&&
 1 +
\frac{1}{4} \int_0^{\infty} dt w_D(t) 
{\overline \acal}_1(t Q^2 e^{-5/3}),
\label{DVsl2}
\\
\lefteqn{
R_{\tau,V}(\triangle S\!=\!0)^{\rm (l.s.)} =
}
\nonumber\\
&&
1 +
\frac{1}{4} \int_0^{\infty} dt W_{\tau}(t) 
{\overline \agoth}_1(t m_{\tau}^2 e^{-5/3}).
\label{Rtausl2}
\end{eqnarray}

To fix the two parameters $M_r$ and $M_0$, we 
choose as experimental inputs the values
\begin{eqnarray}
D_V(Q\!=\!0.7{\rm GeV}) &=& 1.00 \ ,
\label{DVexp}
\\
R_{\tau, V}(\triangle S\!=\!0) &=& 1.196 \ .
\label{Rtauexp}
\end{eqnarray}
The value (\ref{DVexp}) was obtained in Ref.~\cite{PPR} 
on the basis of the ALEPH measurements of $R_V(s)$ \cite{ALEPH}, 
by applying relation (\ref{DVRV}) and neglecting the
light quark masses.
The value (\ref{Rtauexp}) is based on the
measurements of the ALEPH and OPAL Collaborations
\cite{ALEPH}, but with (small) quark mass effects
subtracted out (cf.~also \cite{CL}).
Requiring that the leading skeleton expressions
(\ref{DVsl2}) and (\ref{Rtausl2}) reproduce the
values (\ref{DVexp}) and (\ref{Rtauexp}), respectively,
gives values $M_r = 0.428$ GeV and $M_0 = 0.742$ GeV
in the ${\overline {\rm MS}}$ scheme (this corresponds in
the $V$ scheme to $M_r = 0.985$ GeV and $M_0 = 1.708$ GeV).
With these values of
the scale parameters, we present in
Fig.~1 the curve for $D_V(Q^2)$ as function of $Q$
as predicted by the model, in the leading skeleton
approximation (\ref{DVsl2}), and the ``experimental''
curve \cite{PPR} as obtained by the relation (\ref{DVRV})
(in the massless quark limit)
on the basis of the measured values of $R_V(s)$. 
We see that the two curves agree reasonably well in
the range $Q \stackrel{<}{\sim} 1$ GeV. At higher
values of $Q$, the effects of the $c$ quark mass
become important \cite{Eidelman}. In our calculation 
we used for $\rho_1(\sigma) \equiv
{\rm Im} a_{\rm pt}(-\sigma - i \epsilon)$ the expansion
(\ref{aptexp}) at the three-loop level [i.e., including
terms $\sim\!\ln^3(Q/\Lambda)$.]
 \begin{figure}[th] \unitlength=1mm
\epsfig{file=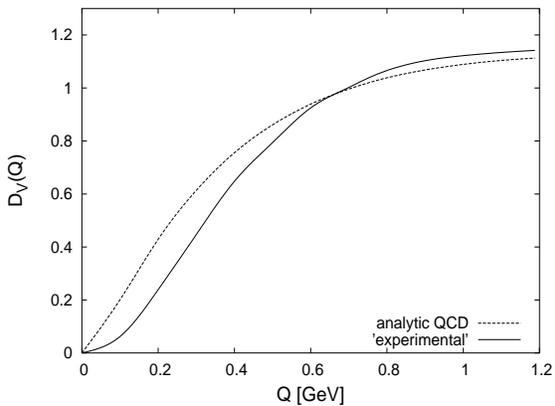,width=7.5cm,height=5.5cm}
\vspace{-1.4cm}
 \caption{\sl\footnotesize  The ``experimental'' curve
for $D_V(Q)$ \cite{PPR}, and the prediction of
our simple model of VMD-motivated analytic QCD in the
leading skeleton approximation.}
\label{Dcanls}
\vspace{-0.6cm}
 \end{figure}

We see that the model, even in the leading
skeleton approximation and in
one of its simplest forms as presented here, 
reproduces reasonably well
the low energy behavior of the Adler function,
and the measured semihadronic $\tau$ decay ratio $R_{\tau}$, while
at the same time the quark masses are set equal
zero. We recall that the minimal
analytic QCD requires large
effective quark masses to reproduce the correct
value of $R_{\tau}$ \cite{MSS}.

In general, at low energies the quark mass effects cannot be
entirely neglected. Further, chiral symmetry
breaking, responsible among other things for a
nonzero pion mass, brings important contributions
to some low energy QCD observables, such as the
axial channel $R_A(s)$ and $D_A(Q^2)$. Such effects
will eventually have to be taken into account
in the framework of the models suggested here.
Furthermore, the contributions beyond the leading skeleton
have to be considered, possibly by analytization
of the power terms $(a_{\rm pt}(Q^2))^n$ ($n \geq 2$)
in the spirit of the approach of \cite{ShS}.
The type of models suggested here can be implemented in
a more sophisticated way than the simplest version
presented here -- e.g., by going beyond
the simple VMD-motivated delta ansatz for the
low energy coupling ${\overline \agoth}_1(s)$.

\end{document}